\documentclass[12pt]{iopart}
\usepackage{iopams}
\usepackage[utf8]{inputenc}
\usepackage{amssymb}
\usepackage{bm}
\usepackage{multirow}
\usepackage{graphicx}
\usepackage[space]{grffile}
\usepackage{xspace}
\usepackage{color}
\usepackage[center]{subfigure}
\usepackage[normalem]{ulem}

\newcommand{\Ms}{\mathrm{M_s}}
\newcommand{\Ma}{\mathrm{M_a}}
\renewcommand{\Re}{\mathrm{Re}}
\newcommand{\Rm}{\mathrm{Rm}}
\newcommand{\Pm}{\mathrm{Pm}}
\newcommand{\apost}{\textit{a posteriori}\xspace}
\newcommand{\aprio}{\textit{a priori}\xspace}
\newcommand{\pd}[2]{\frac{\partial #1}{\partial #2}}
\newcommand{\flt}[1]{\overline{#1}}
\newcommand{\fav}[1]{\widetilde{#1}}
\newcommand{\figref}[1]{\fref{#1}}
\newcommand{\Enzo}{\textsc{Enzo}\xspace}
\newcommand{\FLASH}{\textsc{FLASHv4}\xspace}
\renewcommand{\v}[1]{\boldsymbol{ #1}}
\newcommand{\bra}[1]{\left( #1 \right)}
\newcommand{\grad}[1]{\nabla \bra{#1}}
\newcommand{\curl}[1]{\nabla \times #1}
	
\renewcommand{\div}[1]{\nabla \cdot \bra{#1}}
\newcommand{\emfdata}{{\mathcal{E}}}
\newcommand{\EMFdata}{{\bm{\mathcal{E}}}}
\newcommand{\rres}{\overline{\rho}}

\newcommand{\pres}{\overline{P}}
\newcommand{\emf}{\widehat{\mathcal{E}}}
\newcommand{\EMF}{\widehat{\bm{\mathcal{E}}}}
\newcommand{\tuijdata}[1][]{\tau_{ij}^{\mathrm{u} #1}}
\newcommand{\tbijdata}[1][]{\tau_{ij}^{\mathrm{b} #1}}
\newcommand{\tudata}[1][]{\tau^{\mathrm{u} #1}}
\newcommand{\tbdata}[1][]{\tau^{\mathrm{b} #1}}
\newcommand{\Ekinsgsdata}{E^{\mathrm{u}}_{\mathrm{sgs}}}
\newcommand{\Emagsgsdata}{E^{\mathrm{b}}_{\mathrm{sgs}}}
\newcommand{\Esgsdata}{E_{\mathrm{sgs}}}
\newcommand{\Ekinsgs}{\widehat{E}^{\mathrm{u}}_{\mathrm{sgs}}}
\newcommand{\Emagsgs}{\widehat{E}^{\mathrm{b}}_{\mathrm{sgs}}}
\newcommand{\Esgs}{\widehat{E}_{\mathrm{sgs}}}
\newcommand{\tu}[1][]{\widehat{\tau}_{ij}^{\mathrm{u} #1}}
\newcommand{\tb}[1][]{\widehat{\tau}_{ij}^{\mathrm{b} #1}}
\newcommand{\tunoij}[1][]{\widehat{\tau}^{\mathrm{u} #1}}
\newcommand{\tbnoij}[1][]{\widehat{\tau}^{\mathrm{b} #1}}
\newcommand{\Sij}{\mathcal{S}_{ij}}
\newcommand{\Sijstar}{\mathcal{S}_{ij}^*}
\newcommand{\Mij}{\mathcal{M}_{ij}}
\newcommand{\Cunu}{C^\mathrm{u}_\nu}
\newcommand{\Cbnu}{C^\mathrm{b}_\nu}
\newcommand{\CuE}{C^\mathrm{u}_\mathrm{E}}
\newcommand{\CbE}{C^\mathrm{b}_\mathrm{E}}
\newcommand{\Cunl}{C^\mathrm{u}_\mathrm{nl}}
\newcommand{\Cbnl}{C^\mathrm{b}_\mathrm{nl}}
\newcommand{\Cemf}{C^\mathcal{E}_\mathrm{nl}}
\newcommand{\Ca}{C^\mathcal{E}_{\alpha}}
\newcommand{\Cb}{C^\mathcal{E}_{\beta}}
\newcommand{\Cg}{C^\mathcal{E}_{\gamma}}
\newcommand{\tturb}{t_{\mathrm{turb}}}

\begin{document}

\title[Nonlinear closures for scale separation in supersonic MHD turbulence ]{Nonlinear closures for scale separation in supersonic magnetohydrodynamic turbulence}
\author{Philipp Grete\footnote{corresponding author}$^{1,2}$,
	Dimitar G Vlaykov$^2$,
	Wolfram Schmidt$^2$,
	Dominik R G Schleicher$^2$,
	Christoph Federrath$^3$}
\address{$^1$ Max-Planck-Institut für Sonnensystemforschung, Justus-von-Liebig-Weg 3,
D-37077 Göttingen, Germany}
\address{$^2$ Institut für Astrophysik, Universität Göttingen,
  Friedrich-Hund-Platz 1, D-37077 Göttingen, Germany}
\address{$^3$ Research School of Astronomy \& Astrophysics, 
The Australian National University, Canberra, ACT 2611, Australia}
\ead{\footnotesize{grete@mps.mpg.de}}

\begin{abstract}
Turbulence in compressible plasma plays a key role in many areas
  of astrophysics and engineering.
  The extreme plasma parameters in these environments, e.g. high Reynolds numbers,
  supersonic and super-Alfvenic flows, however, make direct numerical simulations
  computationally intractable
  even for the simplest treatment -- magnetohydrodynamics (MHD).
  To overcome this problem one can use subgrid-scale (SGS) closures -- models for the 
  influence of unresolved, subgrid-scales on the resolved ones.
  In this work we propose and validate a set of constant coefficient closures
  for the resolved, compressible, ideal MHD equations.
  The subgrid-scale energies are modeled by Smagorinsky-like equilibrium 
  closures. The turbulent stresses and the electromotive force (EMF) are described by expressions that 
  are nonlinear in terms of large scale velocity and magnetic field gradients.
  To verify the closures we conduct \aprio tests over 137 simulation snapshots
  from two different codes with
  varying ratios of thermal to magnetic pressure 
  ($\beta_\mathrm{p} = 0.25, 1, 2.5, 5, 25$) and 
  sonic Mach numbers ($\Ms = 2, 2.5, 4$).
  Furthermore, we make a comparison to traditional, phenomenological 
  eddy-viscosity and $\alpha-\beta-\gamma$ closures.
  We find only mediocre performance of the kinetic eddy-viscosity and $\alpha-\beta-\gamma$
  closures, and that the magnetic eddy-viscosity closure
  is poorly correlated with the simulation data.
  Moreover, three of five coefficients of the traditional closures
  exhibit a significant spread in values. 
  In contrast, our new
  closures demonstrate consistently high correlation
  and constant coefficient values over time and and over the wide range of
  parameters tested. 
  Important aspects in compressible MHD turbulence such as the bi-directional energy
  cascade, turbulent magnetic pressure and proper alignment of the EMF
  are well described by our new closures.
\end{abstract}

\noindent {\it Magnetohydrodynamics, Scale separation, Subgrid-scale closure, Turbulence \/}
\date{\today}

\maketitle

\section{Introduction}

Turbulence is ubiquitous in astrophysical plasmas, ranging
from coronal mass ejections and 
stellar winds \cite{1995ARA&A..33..283G},
through star formation in molecular clouds \cite{2004RvMP...76..125M},
to the gas in the interstellar \cite{2004ARA&A..42..211E} and 
intracluster medium.
While experimental setups \cite{2014PhPl...21a3505C} become increasingly 
more realistic, they are
still far away from the regime acting in such extreme conditions.
For numerical simulations it is computationally too expensive 
(if even possible) to capture 
the entire range of physical processes from plasma kinetics to 
the integral scales of turbulence.
In an astrophysical context, one has to further contend 
with the additional complications brought about by high compressibility
and the accompanying supersonic and super-Alfvenic motion.

Possible ways to circumvent the infeasibility of direct numerical simulations
are the
use of calculations based on mean-field theories or
large-eddy simulations \cite{2014arXiv1404.2483S}.
These simulations only resolve the energy containing large scale dynamics and
require a subgrid-scale (SGS) model to account for unresolved effects.
While a lot of research has been successfully carried out in the realm of 
hydrodynamics \cite{2009lesc.book.....G}, compressible magnetohydrodynamic SGS 
closures are essentially unexplored.
Previous research is mainly based on the concept of
turbulent dissipation in incompressible flows
\cite{Mueller2002,1994PhPl....1.3016T,Chernyshov2014,Miki2008}.
They expand the idea of a turbulent eddy-viscosity to an additional 
eddy-resistivity in the induction equation and propose different phenomenological
models.
Even though these models are then evaluated \apost, a general verification and 
justification \aprio has so far only been considered 
for a single incompressible dataset \cite{2010arXiv1010.5759B}. 
Thus, our objective is to establish the validity of closures for the filtered,
compressible MHD equations by coarse-graining multiple datasets from 
high-resolution simulations of statistically homogeneous, forced MHD turbulence.

In general, the effect of finite resolution in numerical simulations can be 
mimicked by applying a low-pass filter to the standard, ideal MHD equations.
This is achieved by convolving the equations with a suitable filter kernel $G$.
See e.g. Garnier \etal \cite{2009lesc.book.....G} for details on the properties of low-pass
filtering and the conditions that $G$ needs to satisfy.
For a homogeneous, isotropic, stationary kernel, under periodic boundary 
conditions \cite{1983PhFl...26.2851F} the equations take the following form
\begin{eqnarray*}
  \pd{\rres}{t}+\div {\rres \fav{\v{u}}} = 0,\\
  \pd{\flt{\rho} \fav{\v{u}}}{t} 
  + \div{\flt{\rho} \fav{\v{u}} \otimes \fav{\v{u}} 
    - \flt{\v{B}} \otimes \flt{\v{B}}}
    + \grad{\pres + \frac{\flt{B}^2}{2}} =&- \nabla \cdot \tau,\\ 
    \pd{\flt{\v{B}}}{t} - \curl\bra{\fav{\v{u}} \times \flt{\v{B}}} =
    \curl{\EMFdata}.
\end{eqnarray*}
The units of the magnetic field $\flt{\v{B}}$ incorporate $1/\sqrt{4\pi}$.
An overbar $\flt{\Box}$ denotes\footnote{Throughout the paper the symbol $\protect\Box$ is 
used as a generic placeholder for variables.
$\protect\widehat{\protect\Box}$ designates closure expressions. Furthermore,
we employ Einstein summation convention and 
$\protect\Box_{i,k}$ is identified with the $k$-th partial derivative of the
$i$-th component of $\protect\Box$.}
filtered and a tilde $\fav{\Box}$ mass-weighted filtered 
quantities \cite{1983PhFl...26.2851F}.
For instance, the filtered density field is given by $\rres=G\ast\rho$, 
while the mass-weighed filtered velocity field is $\fav{\v{u}} = \flt{\v{u}\rho}/\rres$. 
In this formalism, all filtered primary quantities, density $\rres$, 
velocity $\fav{\v{u}}$, magnetic field $\flt{\v{B}}$, and thermal pressure $\pres$
are presumed to be known and directly accessible.
Due to the introduction of mass-weighted filtering the only remaining terms 
that require closure are the SGS stress $\tau$ 	
and the electromotive force (EMF), $\EMFdata$.
They are analytically expressed \cite{Miki2008} as
\begin{eqnarray}
  \EMFdata &= \flt{\v{u}\times\v{B}} - \fav{\v{u}} \times
\flt{\v{B}}\label{eq:EMFdata}
  \quad \mathrm{and} \\
  \label{eq:tau_def}\tau_{ij} &=  \tuijdata - \tbijdata +
  \bra{\flt{B^2} - \flt{B}^2}\frac{\delta_{ij}}{2}, \quad \mathrm{with} \nonumber \\
  \tuijdata &\equiv \flt{\rho} \bra{\fav{u_i u_j} - \fav{u}_i \fav{u}_j}
  \qquad \mathrm{and} \qquad
  \tbijdata \equiv \bra{\flt{B_i B_j} - \flt{B}_i~\flt{B}_j}\;. \label{eq:tudata}
\end{eqnarray}
The SGS stress tensor can be decomposed into the well-known turbulent 
Reynolds stress $\tau^{\mathrm{u}}$,
a turbulent Maxwell stress $\tau^{\mathrm{b}}$ and 
a magnetic pressure term.

Furthermore, the definitions of the SGS energies are obtained from applying the
filter to the total filtered energy density $\flt{E}$, which can be decomposed into
the contribution due to resolved fields only and a remainder, designated as SGS energy
\begin{eqnarray*}
	\flt{E} = \underbrace{\frac{1}{2} \flt{\rho} \fav{u}^2 + \frac{1}{2}\flt{B}^2}_
	{\mathrm{(resolved)}}
	+ \underbrace{\frac{1}{2}\flt{\rho} \bra{\fav{u^2} - \fav{u}^2} + \frac{1}{2} \bra{\flt{B^2} - \flt{B}^2}}_
	{= \Ekinsgsdata + \Emagsgsdata \equiv \Esgsdata \; \mathrm{(unresolved)}}.
\end{eqnarray*}

It is important to point out that in general the filtering operator is not a Reynolds operator, 
in particular $\flt{\flt{\Box}}\neq\flt{\Box}$. 
It follows that SGS terms, like $E_{SGS}$, carry information not only about the interactions
between unresolved fields but also about cross-scale interactions between unresolved and resolved fields.
In addition to this, 
the turbulent magnetic pressure is identical to the magnetic 
SGS energy $\Emagsgsdata$
and both kinetic and magnetic SGS energies are
directly given by $2\Ekinsgsdata = \Tr\bra{\tau^{\mathrm{u}}}$ and
$2\Emagsgsdata = \Tr\bra{\tau^{\mathrm{b}}}$,
i.e. they constitute the isotropic parts of the respective SGS tensors.
Following the general tensor decomposition, the deviatoric, traceless
parts are then
given by $\tau^{\Box *}_{ij} = \tau^{\Box}_{ij} - \frac{1}{3}\delta_{ij}\tau^{\Box}_{kk}$.
\section{Traditional closures}
\label{sec:TradClosures}

In hydrodynamics, the traceless part of the SGS stress tensor 
is commonly closed
by means of the eddy-viscosity hypothesis 
$\tu[*] =  - 2 \nu^\mathrm{u} \flt{\rho} \fav{\Sijstar}$
in analogy to the molecular viscosity term in the momentum equation, where
$\fav{\Sij} \equiv \frac{1}{2}\bra{\fav{u}_{i,j} + \fav{u}_{j,i}}$
is the filtered kinetic rate-of-strain tensor.
This introduces a turbulent kinetic eddy-viscosity 
$\nu^\mathrm{u}~=~\Cunu \Delta \bra{\Ekinsgsdata/\flt{\rho}}^{1/2}$ which is proportional
to a characteristic velocity, commonly given by the kinetic SGS energy, 
and a characteristic length scale $\Delta$.
This closure has already been applied directly to MHD 
\cite{Mueller2002,Petro2007} by neglecting the magnetic contribution
$\tb[*]$ in the momentum equation.
A turbulent magnetic viscosity  
$\nu^\mathrm{b} = \Cbnu \Delta \bra{\Emagsgsdata}^{1/2}$ was used
in \cite{Miki2008} with the closure
\begin{eqnarray*}
  \tb[*] =  - 2 \nu^\mathrm{b} \flt{\Mij} \;,
\end{eqnarray*}
where $\flt{\Mij} \equiv \frac{1}{2}\bra{\flt{B}_{i,j} + \flt{B}_{j,i}}$ 
is the filtered magnetic rate-of-strain tensor.

The SGS energies can either be determined by individual evolution equations,
where several terms again require closure, or by an instantaneous closure.
Smagorinsky \cite{Smagorinsky1963} introduced such an instantaneous closure
in pure incompressible hydrodynamics ($\v{B} = 0$) by assuming the SGS energy 
flux to be in equilibrium
with the rate of dissipation
\begin{eqnarray}
  \label{eq:SmagU}
  \Ekinsgs=  \CuE \Delta^2 \rho |\fav{\mathcal{S}}^*|^2 \;.
\end{eqnarray}
Here, $|\fav{\mathcal{S}}^*| \equiv \sqrt{2 \fav{\Sijstar} \fav{\Sijstar}}$ 
denotes the rate-of-strain magnitude. 

Finally, the EMF is commonly modeled 
(e.g. \cite{Mueller2002,1994PhPl....1.3016T,Petro2007,Miki2008})
by variations of 
\cite{Yoshizawa1990}
\begin{eqnarray*}
  \EMF = \alpha \flt{\v{B}} - \beta \flt{\v{J}} + \gamma \fav{\v{\Omega}},
\end{eqnarray*}
with resolved
current $\flt{\v{J}}=\curl \flt{\v{B}}$ and
vorticity $\fav{\v{\Omega}} = \curl \fav{\v{u}}$.
The coefficients  $\alpha$, $\beta$, and $\gamma$ are typically related
to the $\alpha$-effect, turbulent resistivity and turbulent cross helicity,
respectively.
The commonly used closures for these coefficients,
\begin{eqnarray*}
  \alpha = \Ca \tturb H, \qquad &
  \beta = \Cb \tturb \Esgsdata/\flt{\rho}, \qquad	 &
  \gamma = \Cg \tturb W, \qquad
\end{eqnarray*}
 are based on dimensional arguments, with turbulent
cross helicity $W = \flt{\v{u}\cdot \v{B}} - \fav{\v{u}}\cdot \flt{\v{B}}$,
residual helicity 
$H \sim \bra{\flt{\v{J} \cdot \v{B}}-\flt{\v{J}} \cdot \flt{\v{B}}} - \flt{\rho}\bra{\flt{\v{\Omega} \cdot \v{u}}-\fav{\v{\Omega}} \cdot 
\fav{\v{u}}}$,
and timescale $\tturb~=~\Delta~\bra{\Esgsdata/\rres}^{-1/2}$.

\section{Nonlinear closures}

\label{sec:NonLinClosures}
In our new approach we adopt the compressible hydrodynamic nonlinear closure for the kinematic deviatoric 
stress tensor $\tudata[*]$
from \cite{Schmidt2011}, similar to the incompressible one from~\cite{Woodward2006}.
We propose the straightforward extension to MHD with 
\begin{eqnarray}
  \label{eq:tuStar}
  \tu[*] &= 2 \Cunl \Ekinsgsdata \left (
  \frac{\fav{u}_{i,k} \fav{u}_{j,k}}{\fav{u}_{l,s} \fav{u}_{l,s}} 
  - \frac{1}{3} \delta_{ij} \right), \\
  \label{eq:tbStar}
  \tb[*] &= 2 \Cbnl \Emagsgsdata \left ( 
  \frac{\flt{B}_{i,k} \flt{B}_{j,k}}{\flt{B}_{l,s} \flt{B}_{l,s	}} 
  - \frac{1}{3} \delta_{ij} \right) \;.
\end{eqnarray}

The tensorial structure, e.g. $\fav{u}_{i,k} \fav{u}_{j,k}$,
can be obtained by a Taylor expansion 
discarding terms with $2^\mathrm{nd}$ and higher order gradients of the resolved fields.
The overall normalisation with the subgrid-scale energies comes from the constraint that the SGS stresses
vanish in laminar flows.

Applying the nonlinearity idea to the EMF
generalizes the closure proposed by \cite{2010arXiv1010.5759B} to 
the compressible
regime
\begin{eqnarray}
  \emf_{i} = \varepsilon_{ijk} \Cemf \Delta^2 \fav{u}_{j,s} \flt{B}_{k,s} \;.
\end{eqnarray}
The closure explicitly preserves the anti-symmetry 
between velocity and magnetic field in $\EMFdata$, which in turn helps in
capturing their relative geometry.

Finally, to complete the set of nonlinear closure equations, 
we use the Smagorinsky expression
for the turbulent kinetic energy  \eref{eq:SmagU} 
and propose an analogous extension to the magnetic part
\begin{eqnarray}
  \label{eq:SmagB}
\Emagsgs = \CbE \Delta^2 |\flt{\mathcal{M}}|^2 \;.
\end{eqnarray}
Here, the turbulent magnetic energy 
is proportional to the magnetic rate-of-strain magnitude 
$|\flt{\mathcal{M}}| \equiv \sqrt{2 \flt{\Mij}\, \flt{\Mij}}$.
There are two advantages of closing $\Ekinsgsdata$ and $\Emagsgsdata$ 
separately and not jointly via the total SGS energy.
First, there is no additional need to close the often neglected turbulent 
magnetic pressure, as it is given by $\Emagsgsdata$.
Second, the individual energies provide closures to the isotropic parts of
the turbulent stress tensors $\tudata$ and $\tbdata$.

\section{Validation Method}
\label{sec:Validation}
In order to evaluate 
the proposed closures, we perform an \aprio 
comparison using simulation data obtained
from two, grid-based MHD codes (\Enzo \cite{Enzo2013} and \FLASH~\cite{Fryxell2000}). 
This way the results are less likely to hinge on the particulars of the numerical implementation.
In both cases we follow the evolution of a compressible, isothermal fluid in
a cubic box with resolution of $512^3$ grid cells and periodic boundary 
conditions, starting from uniform initial conditions. 
In \Enzo we use an ideal equation of state with adiabatic exponent 
$\kappa = 1.001$ in order to approximate isothermal gas.
 \Enzo is a finite-volume code, i.e. the evolution equations are evaluated in integral form by solving
a Riemann problem for the mass, momentum and energy flux through cell walls. This allows for the conservation of MHD invariants  (e.g. energy) 
to machine precision.
We use a MUSCL-Hancock scheme \cite{Waagan2011} (a second-order accurate Godunov extension) with second-order Runge-Kutta time integration and
Harten-Lax-van Leer (HLL) Riemann solver (a two-wave, three-state solver) to solve the ideal MHD equations. 
The \FLASH code is similar to \Enzo (second-order accurate in space and time), but uses the positive-definite
HLL3R Riemann solver \cite{Waagan2011}. Another difference is that \FLASH uses a polytropic equation of state to keep the gas
exactly isothermal. Moreover, explicit kinematic viscosity and magnetic resistivity terms are included in the momentum,
energy, and induction equations. We set the kinematic and magnetic Reynolds
numbers to $\Re=\Rm=3780$.
Consequently, the magnetic Prandtl number $\Pm=\Rm/\Re$ is unity. For details on the numerical methods used in \FLASH, 
including viscous and resistive dissipation, see \cite{Federrath2011} and \cite{Federrath2014}.
Both codes employ divergence cleaning \cite{Dedner2002} to maintain 
$\nabla \cdot \v{B} = 0$.
A state of homogeneous and isotropic turbulence is reached by supersonic 
stochastic driving in the momentum equation (given by an Ornstein-Uhlenbeck process) at small
wave-numbers, similar to \cite{Schmidt2009,2010A&A...512A..81F}.
Thus, the forcing field is evolving in time and space.
The associated large auto-correlation time-scale $T$ of the forcing 
translates to the eddy turnover time of the largest, energy-containing eddies.
It is therefore the chosen unit of time in the following.

We explore a range of parameters. 
The initial strength of the magnetic field is set by the plasma $\beta_\mathrm{p}$
-- the ratio of thermal to magnetic pressure.
The final sonic Mach number $\Ms$ is determined by the forcing amplitude.
For the \Enzo simulations we have initial $\beta_\mathrm{p} = 0.25, 2.5, 25$, 
with $\Ms \approx 2.5$ after $t \approx 2T$ turnover times. 
The \FLASH simulations reach $\Ms \approx 4, 2$ for initial $\beta_\mathrm{p} = 1, 5$, 
keeping instead constant Alfvenic Mach number $\Ma \approx 3$.
We discard all initial data affected by transients (before a simulation time 
of $t=2T$) and analyze consequent snapshots taken approximately in intervals of
$0.15T$ and $0.1T$ for the \Enzo and \FLASH datasets, respectively.

The analysis begins with the application of a low-pass Gaussian (test) filter
to the equations of motion.
In the context of the closures we investigate, filtered quantities
(i.e. density, velocity, and magnetic field)
are interpreted as resolved, while the remainders represent the 
unresolved small scales.
We can then compute $\tau$ and $\EMFdata$ both
directly from \eref{eq:EMFdata} and \eref{eq:tudata},
and from their respective traditional and nonlinear closures.

\begin{figure}[!ht]
	\centering
\hfill
	\subfigure[kinetic energy\label{fig:kin spectrum}]{
		\includegraphics[width=0.4\textwidth]{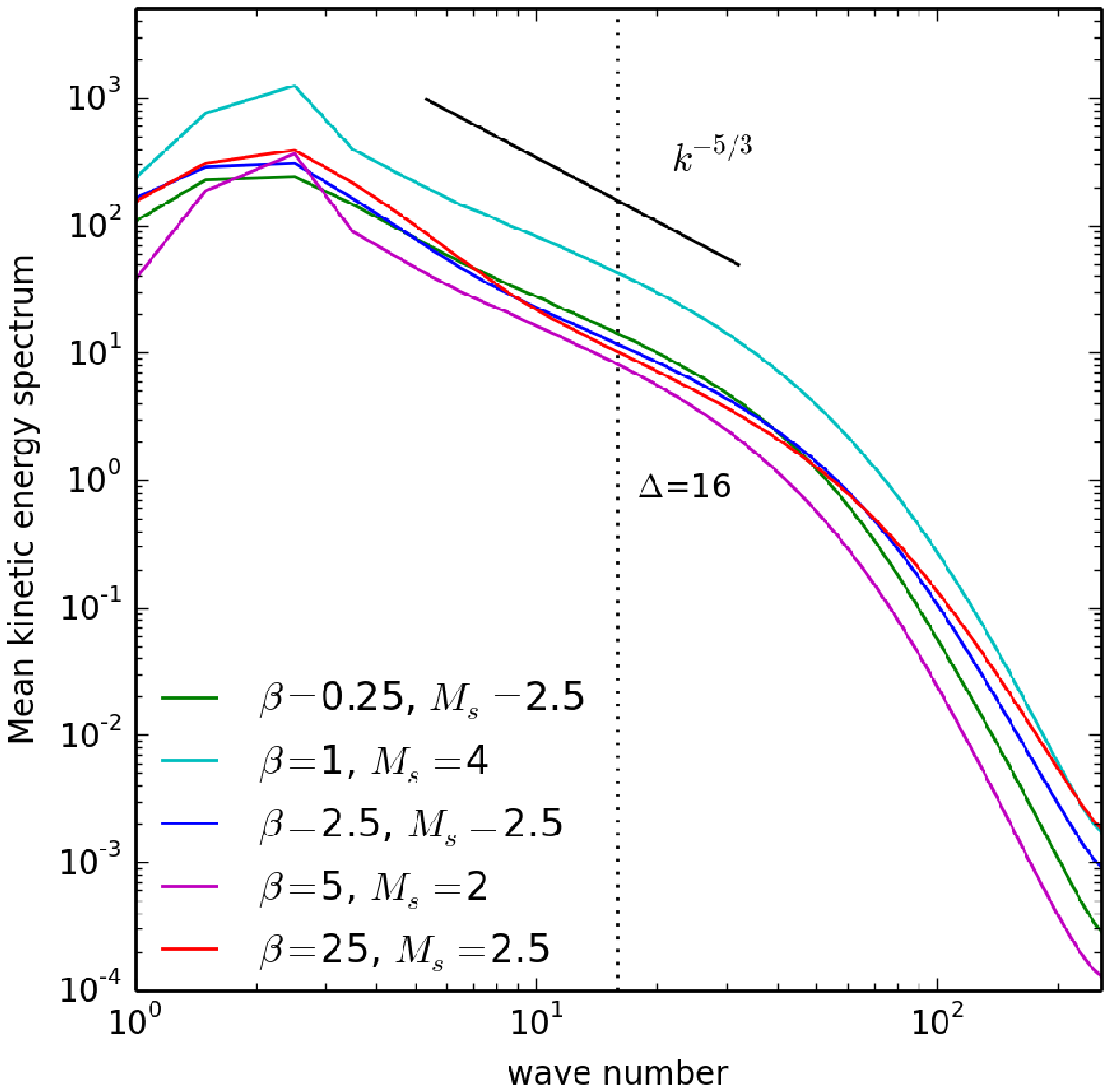}
    }
    \hfill
    \subfigure[kinetic $+$ magnetic energy\label{fig:tot spectrum}]{
		\includegraphics[width=0.4\textwidth]{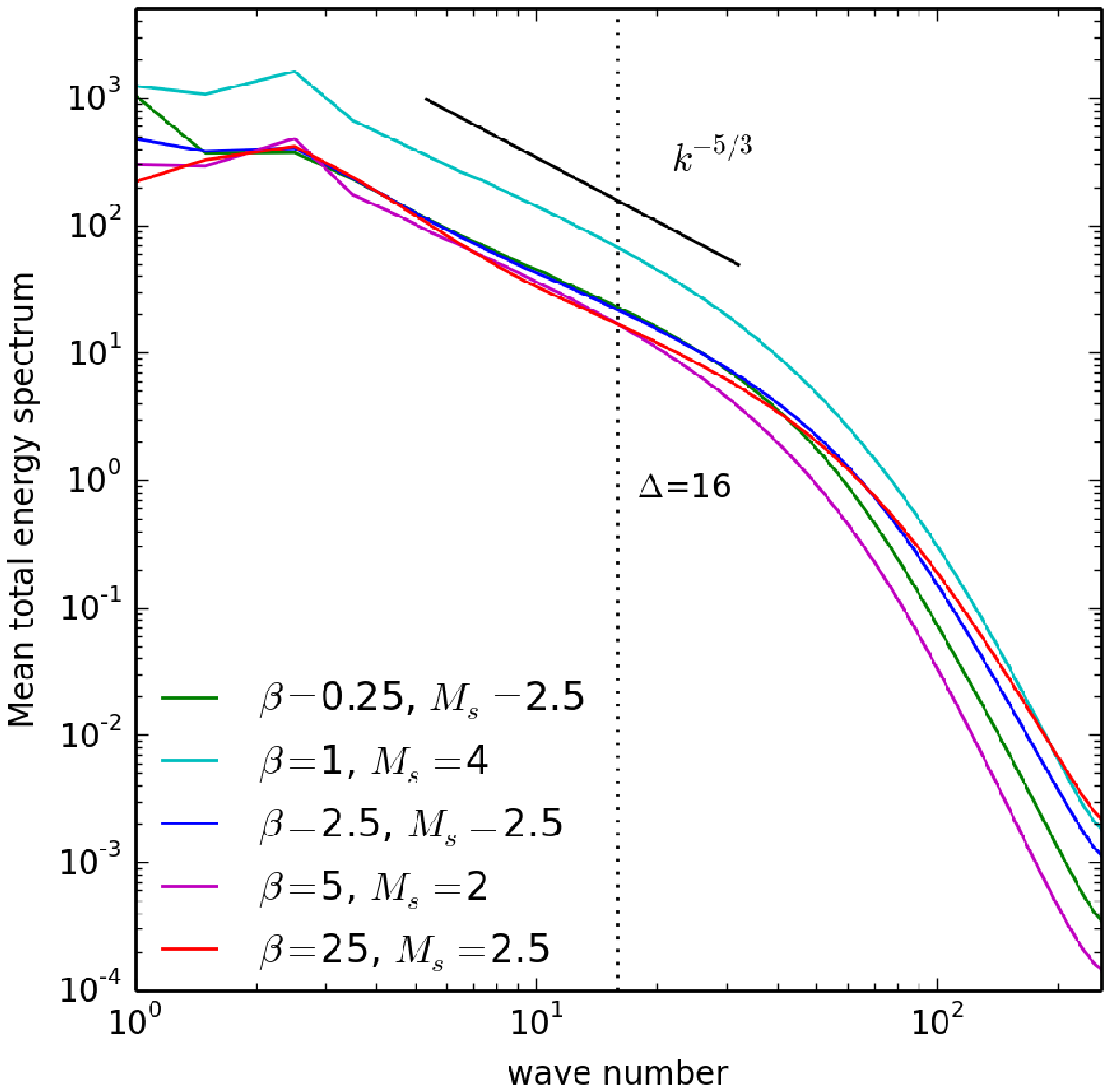}
    }
    \hfill
  \caption{
  Kinetic (a) and total (b) energy spectrum for each dataset, averaged over the time between $2T$ and $5T$.
  The kinetic energy is calculated from the Fourier transform of $\sqrt{\rho}\v{u}$.
  }
  \vspace{-1em}
  \label{fig:spectrum}
\end{figure}

The determination of the length-scale of the filter bears some consideration.
It needs to fall within an intermediate range of length scales, away from
the particular effects of both the large-scale forcing and the small-scale dissipation.
The largest scale of the system is the full box size $L$ and corresponds to the wavenumber
$n=1$, while the smallest scale is given by the Nyquist wavenumber $n_\mathrm{Nyq}=N/2=256$ for the linear
numerical resolution $N=512$ grid cells. The turbulence injection wavenumber is $n_\mathrm{inj} = 2$ 
(corresponding to half the box size $L/2$) in both codes, which is why the
energy spectra, 
as illustrated in \fref{fig:spectrum}, peak there.
The figure shows the mean kinetic and total (kinetic plus magnetic) energy 
spectra as a function of wavenumber.
Since the stochastic forcing is implemented only in the momentum equation both for \Enzo and \FLASH, 
the kinetic energy spectrum exhibits the most direct imprint of the forcing itself. Conversely, the total energy
spectrum carries the overall effect of the small-scale dissipation through both kinetic and magnetic channels.
\Fref{fig:spectrum} demonstrates that our simulations produce approximate power-law scaling 
within a narrow range of wavenumbers, which is indicative of self-similar
turbulent fluctuations \cite{Federrath2013a}.
This can be interpreted as inertial range dynamics, although the nature of the inertial range 
in compressible MHD turbulence is still not fully understood
\cite{Galtier2011, Banerjee2013, Aluie2011, Aluie2013}. 
Furthermore, this range separates the forcing scale and the dissipation scales and is not affected by
numerical diffusion in the absence of a bottleneck effect as demonstrated by \cite{Kitsionas2009}.
The vertical dotted line in \fref{fig:spectrum} indicates our chosen filter length scale,
corresponding to $\Delta = 16$ grid cells
or wavenumber $n_\mathrm{filter} =N/(2\Delta) = 16$.
This filter scale falls within the range of
the self-similar power-law range for both the kinetic and total energy spectra.
This is why we use this ideal scale for our filter in the following analysis.

Additionally, this provides the motivation to treat data from both simulations on equal footing,
even though \FLASH has explicit viscosity and diffusivity while \Enzo solves the ideal MHD equation,
subject to numerical dissipation only.

In order to incorporate coordinate independence, a scalar field is chosen for 
comparison, the SGS energy flux $\Sigma$, i.e. the term responsible for
the transfer of SGS energy between resolved and unresolved scales.
Its components associated with the Reynolds and Maxwell stresses and the EMF are
$\Sigma^u=\tau^u_{ij} \fav{S}_{ij}$, $\Sigma^b=\tau^b_{ij} \fav{S}_{ij}$  and
$\Sigma^{\emfdata}=\EMFdata \cdot \flt{\v{J}}$, respectively
(see appendix of \cite{Miki2008} for the detailed SGS energy equation).
Here, we substitute \eref{eq:EMFdata} and \eref{eq:tudata} to obtain
exact fluxes $\Sigma^\Box$ and match these to model fluxes 
$\widehat{\Sigma}^\Box$ that employ the corresponding closures.
For example, in the case of the eddy-viscosity closure for the deviatoric
turbulent Reynolds stress tensor we compare the exact flux
\begin{eqnarray}
	\Sigma^{\mathrm{u*}} =
	\tau_{ij}^{\mathrm{u*}} \fav{\Sij} =
 	\flt{\rho} \bra{ \bra{\fav{u_i u_j} - \fav{u}_i \fav{u}_j}
	- \frac{1}{3} \delta_{ij} 
	\bra{\fav{u_k u_k} - \fav{u}_k \fav{u}_k}}
	\fav{\Sij}
\end{eqnarray}
with the model flux
\begin{eqnarray}
	\widehat{\Sigma}^{\mathrm{u*}} =
	\tu[*] \fav{\Sij} =
	- \nu^\mathrm{u} \flt{\rho} |\fav{\mathcal{S}^*}|^2 
	= - \Cunu \Delta \bra{\flt{\rho} \Ekinsgsdata }^{1/2} 
	|\fav{\mathcal{S}^*}|^2 \;.
	\label{eq:eddyViscFlux}
\end{eqnarray}
On the one hand, the comparison involves the determination of the 
constant (in space and time), dimensionless closure coefficients 
$C_{\Box}^{\Box}$.
They are computed individually for each snapshot 
by minimizing the error between $\Sigma^\Box$ and $\widehat{\Sigma}^\Box$
in the least-square sense.
This allows to further test the constancy of the coefficients with respect to 
time and plasma parameters.
On the other hand, the general performance of the closure is gauged by computing the 
Pearson correlation coefficient of $\Sigma^\Box$ and $\widehat{\Sigma}^\Box$, where 
the obtained closure coefficients are substituted in. 

Several assumptions should be pointed out concerning this validation technique.
Firstly, the simulation data we have available for comparison fall short of realistic
astrophysical parameters, e.g. with regards to Reynolds numbers and resolution.
In that sense, it would be interesting to use higher resolution direct numerical simulation data or
three-dimensional observations or experimental results. The problem is that
experimental data for supersonic compressible turbulent plasmas are not available
and obtaining realistic Reynolds numbers is computationally challenging 
for astrophysical parameters.
However, as seen from \figref{fig:spectrum}, the data we have are sufficiently well resolved for 
our analysis.
Secondly, in choosing the SGS energy flux $\Sigma$, as a diagnostic variable,
we implicitly assume that in the context of homogeneous and isotropic turbulence 
the turbulent transport (encoded by terms of the form 
$ \div {\fav{u}\cdot \tau}$ and $\div{\flt{B}\times \EMFdata}$)
averages out to zero on subgrid scales.
This assumption can nevertheless be easily relaxed by incorporating further diagnostic variables. 
Finally, we have focused on the SGS energy since it increases monotonically with the strength of turbulence
regardless of the type of turbulence (e.g. compressive or solenoidal, weak or strong, etc.).
As an extension, the other two quadratic MHD invariants -- the magnetic helicity and cross-helicity, 
may further highlight distinct turbulence properties present in particular flow configurations.
These should be kept in mind as further avenues of investigation, once a preferred closure 
has been identified by the described validation technique.

\section{Results}
\begin{table}
  \caption{\label{tab:valueoverview}
    Model coefficient overview -- coefficient value and energy flux correlation:
 median and bounds of the central 90\% interval across all datasets.}
\begin{indented}  
\item[]
\lineup
\begin{tabular}{@{}lcll}
  
  \br
  \multicolumn{1}{c}{Model} & Coefficient & Value & Corr[$\Sigma^\Box,\,\widehat{\Sigma}^\Box$]\\
  \mr
  \multirow{2}{*}{Smagorinsky}
  & $\CuE$ & $0.056^{+0.016}_{-0.015}$   & $0.90^{+0.024}_{-0.04}$ \\
  & $\CbE$ & $0.075^{+0.034}_{-0.007}$   & $0.91^{+0.021}_{-0.04}$  \\
  \mr
  \multirow{2}{*}{eddy-viscosity}
  & $\Cunu$ & $0.061^{+0.045}_{-0.019}$  & $0.70^{+0.13}_{-0.11}$ \\
  & $\Cbnu$ & $-0.002^{+0.029}_{-0.03}$ & $0.06^{+0.14}_{-0.06}$    \\
  \mr
  \multirow{3}{*}{nonlinear}
  & $\Cunl$ & $0.68^{+0.09}_{-0.09}$   & $0.94^{+0.04}_{-0.04}$   \\
  & $\Cbnl$ & $0.77^{+0.08}_{-0.12}$   & $0.90^{+0.04}_{-0.07}$   \\
  & $\Cemf$ & $0.12^{+0.013}_{-0.024}$ & $0.79^{+0.07}_{-0.17}$\\
  \mr
  \multirow{3}{*}{$\alpha-\beta-\gamma$}
  & $\Ca$ & $0.0007^{+0.0010}_{-0.0016}$ & \multirow{3}{*}{\Bigg\} $0.58^{+0.06}_{-0.16}$ }\\
  & $\Cb$ & $0.020^{+0.009}_{-0.005}$ & \\
  & $\Cg$ & $-0.005^{+0.067}_{-0.045}$ & \\
  \br
\end{tabular}
\end{indented}
\end{table}

\begin{figure}[!ht]
	\centering
	\hfill
	\subfigure[kinetic SGS energy\label{fig:EsgsU}]{
		\includegraphics[width=0.4\textwidth]{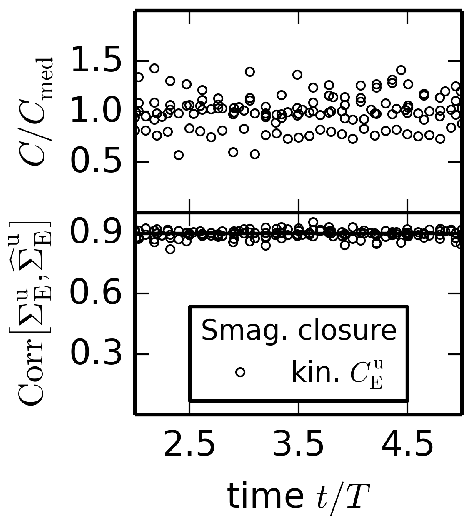}
    }
    \hfill
    \subfigure[magnetic SGS energy\label{fig:EsgsB}]{
		\includegraphics[width=0.4\textwidth]{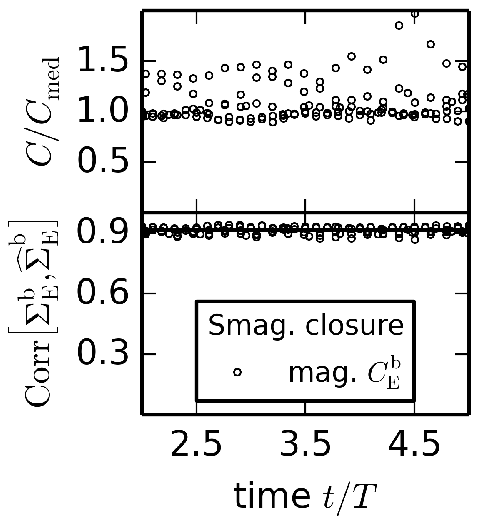}
    }
    \hfill
    \caption{
    Model coefficient values (top panels) and correlations (bottom panels with 
    --- median) from fitting model energy flux 
    $\widehat{\Sigma}^\Box_{\mathrm{E}} = \widehat{\tau}^\Box_{ij}\fav{\Sij}$ 
      to exact flux for the isotropic parts of the SGS stress tensors.
      These are given by the respective energy model in the trace elements
      ($\tau^\Box_{ii} = \frac{2}{3} E^\Box_{\mathrm{sgs}}$). 
      Each panel contains the joint data of all
      simulations and each snapshot is represented by a marker.
      Values are given in \tref{tab:valueoverview}.}
  \label{fig:Esgs}
\end{figure}

\begin{figure}[!ht]
	\centering
	\subfigure[traceless kinetic SGS stress tensor\label{fig:tauUstar}]{
		\includegraphics[width=0.45\textwidth]{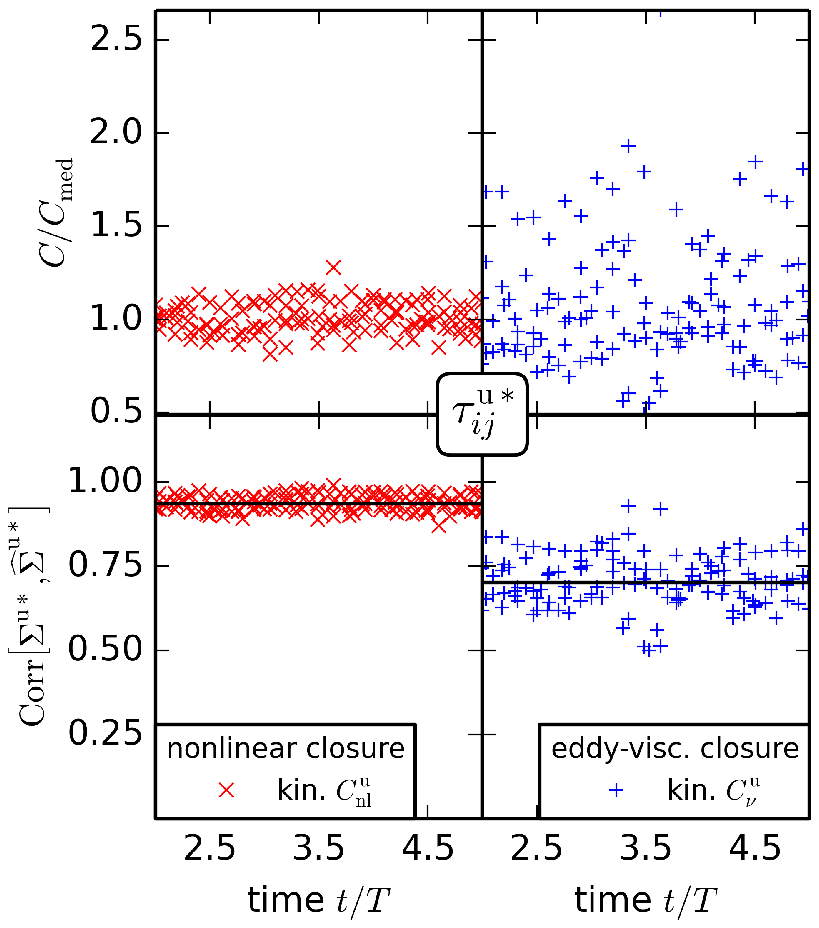}
    }
    \hfill
    \subfigure[traceless magnetic SGS stress tensor\label{fig:tauBstar}]{
		\includegraphics[width=0.49\textwidth]{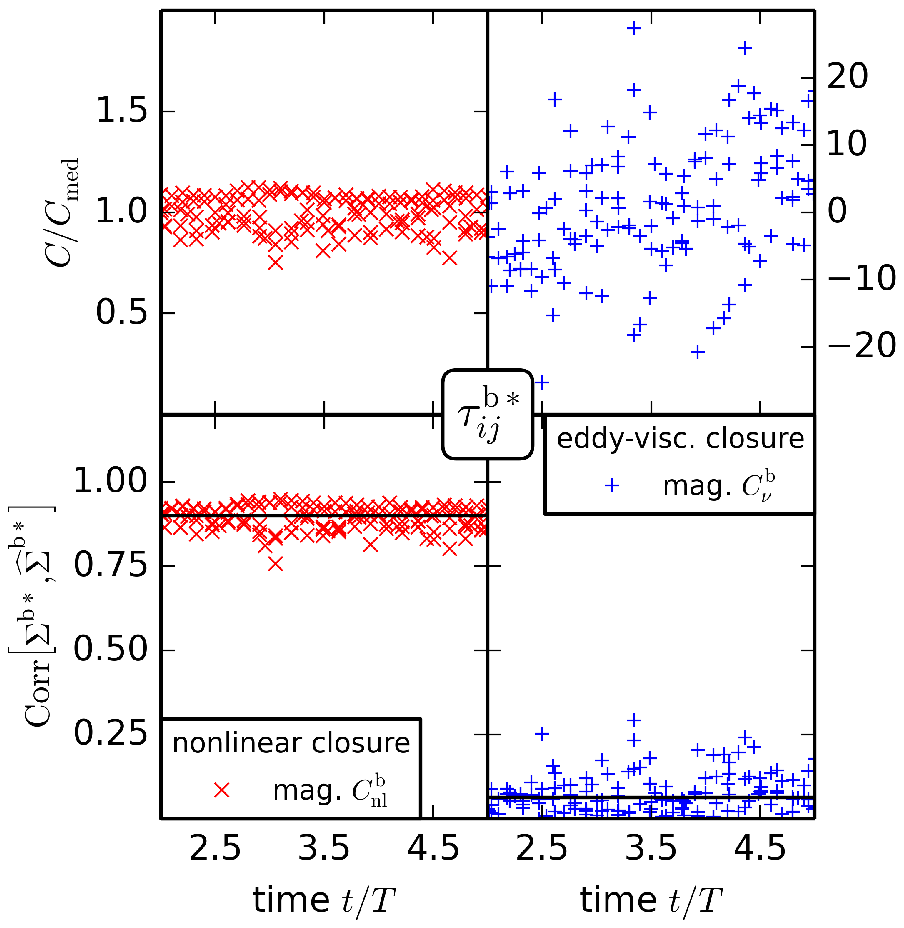}
    }
    \caption{
    Model coefficient values (top panels) and correlations (bottom panels with --- median)
    from fitting model energy flux 
      $\widehat{\Sigma}^\Box = \widehat{\tau}^\Box_{ij}\fav{\Sij}$ 
      to exact flux for the nonlinear closure (left panels) and eddy-viscosity closure (right panels). 
      Each panel contains the joint data of all
      simulations and each snapshot is represented by a marker.
      Values are given in \tref{tab:valueoverview}.}
  \label{fig:tauStar}
\end{figure}

The fitting results for the SGS stress tensors' energy flux
are given in \figref{fig:Esgs} for the isotropic components and \figref{fig:tauStar}
for the deviatoric components.

The isotropic parts of $\tudata$ and $\tbdata$ are given
by the SGS energies $\tau^\Box_{ii} = \frac{2}{3} E^\Box_{\mathrm{sgs}}$ from
\eref{eq:SmagU} and \eref{eq:SmagB}.
Both, the coefficient values of kinetic part (\figref{fig:EsgsU} top panel) 
and the magnetic part (\figref{fig:EsgsB} top panel), have a small spread within
a factor of two across time and all simulations.
Furthermore,  closure and data are highly correlated (bottom panels) with a 
median correlation coefficient of 0.90 and 0.91, respectively.
More detailed numerical values of these and all following coefficients and 
correlations are listed in \tref{tab:valueoverview}.

The differences in the deviatoric parts 
$\tunoij[*]$ in \figref{fig:tauUstar} and $\tbnoij[*]$ in \figref{fig:tauBstar}
between the nonlinear and the eddy-viscosity closures
are apparent.
While our nonlinear closure exhibits approximately constant coefficient values
and correlations over time in all simulations, the 
kinetic eddy-viscosity closure shows a correlation weaker by $\approx 0.2$ and 
	bigger spread in coefficient values.
	Moreover, the magnetic eddy-viscosity closure is effectively uncorrelated with
	the simulation data and
the coefficients can even switch sign at different times.
\begin{figure}[!h]
  \centering
  \includegraphics{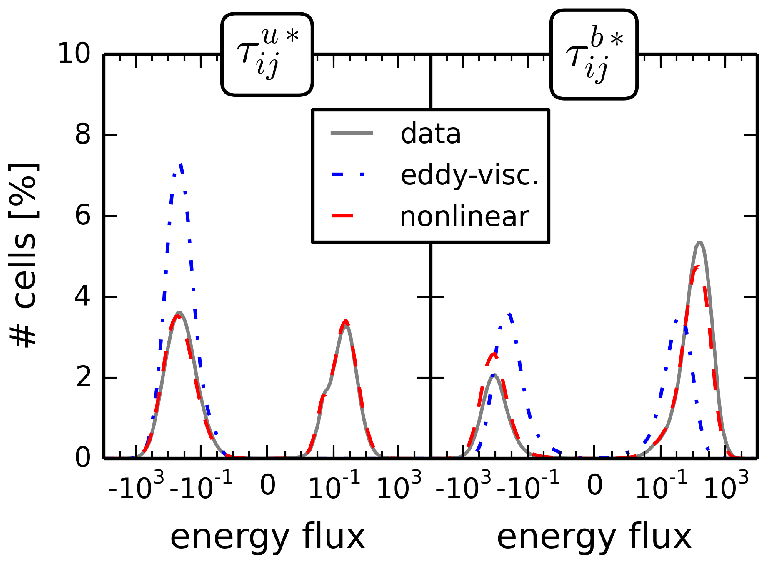}
  \vspace{-1em}
  \caption{
  Representative snapshot (\Enzo sim. with 
  $\beta_\mathrm{p} = 2.5$ at $t = 4.44T$) of the energy flux 
    $\Sigma^\Box = \tau^\Box_{ij}\Sij$
    distribution within the simulation box.
  }
  \vspace{-1em}
  \label{fig:fluxDistribution}
\end{figure}
The performance of the different closures can be understood from
\figref{fig:fluxDistribution}, where we plot the energy flux distributions
$\Sigma^\mathrm{u*}$ and $\Sigma^\mathrm{b*}$ for a single snapshot.
A negative flux $\Sigma^{\mathrm{u}*}<0$ corresponds to a forward energy cascade
--
the transfer of energy from resolved to subgrid scales, because
$\Sigma^{\mathrm{u}*}$ appears
as a sink term in the SGS kinetic and magnetic energy evolution equations and
as a source term in the respective resolved energy equations. 
Conversely, a positive flux corresponds to an inverse energy cascade,
i.e. transport of energy from subgrid to resolved spatial scales.
The general distribution of the actual fluxes in \figref{fig:fluxDistribution}
is representative for all snapshots.
The kinetic SGS energy fluxes are globally almost 1:1 in both directions of the
turbulent cascade with a slight tendency towards the forward cascade.
However, the forward cascade is about 3-10 times stronger depending on the parameters
as indicated by the position of the peaks in the distribution.
For this reason, the kinetic eddy-viscosity closure shows a moderate 
correlation even though it captures only the forward energy cascade --
from large to small scales. 
In fact, since under the eddy-viscosity hypothesis
the kinetic SGS energy flux has the form
$\widehat{\Sigma}^{\mathrm{u*}}=-\nu^\mathrm{u} \flt{\rho}|\fav{S}^*|^2$, 
see \eref{eq:eddyViscFlux}, 
any model in which the eddy-viscosity $\nu^\mathrm{u}$ has a definite signature
with respect to space cannot reproduce a bi-directional energy cascade
that is well represented by the nonlinear closure. 
In contrast to the kinetic SGS energy flux, the global magnetic flux clearly has
a preferred direction.
Depending on the parameters, between 60\% and 80\% of cells have a positive SGS magnetic
flux 
indicating energy transfer from large to small scales.
Nevertheless, the difference in strength is less pronounced as the overall
forward flux is
about 2 times stronger than the backward one.
Again, these properties are well captured by the proposed nonlinear closure whereas
the magnetic eddy-viscosity closure is poorly correlated in both strength and magnitude.
Finally, it should be noted, that the nonlinear closures also work 
very well with the Smagorinsky energy closure.
Exchanging the exact expressions $\Esgsdata^\Box$ in \eref{eq:tuStar} and
\eref{eq:tbStar} with $\Esgs^\Box$ only slightly reduces the correlations
(max. 5\%) and the coefficients remain constant up to the second 
significant figure (not plotted here).

\begin{figure}[ht!]
    \centering
    \includegraphics{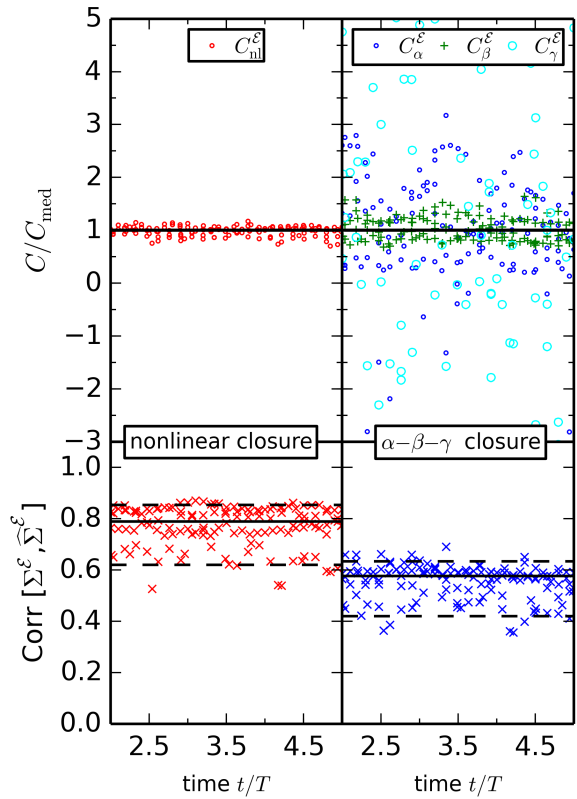}
    \caption{
    Model coefficient values (top panels) normalized to the sample 
      median (---) and the corresponding Pearson
    correlation coefficients (bottom panels) with 90\% central interval (- -) 
    for the nonlinear closure 
    (left panels) and the reference closure (right panels) from 
      energy flux fitting for $\EMF$. Each panel contains the joint data of all
      simulations and each snapshot is represented by a marker.
      Values are listed in \tref{tab:valueoverview}.
    }
  \label{fig:EMF_corr}
\end{figure}

\begin{figure}[ht!]
    \centering
    \includegraphics{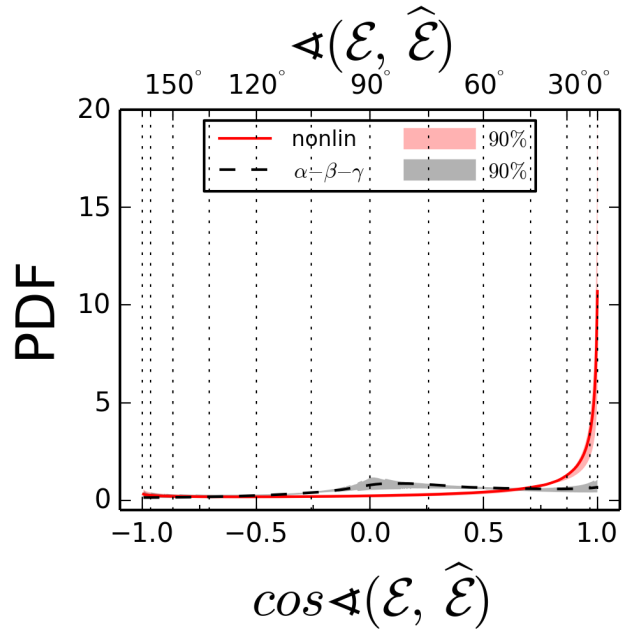}
    \caption{
    Probability density plot of the local (cell-by-cell) alignment between $\EMFdata$ and $\EMF$. The lines designate the median PDF across all times
    and datasets for the nonlinear closure (red ---) and
    the $\alpha-\beta-\gamma$ closure (gray - -).
    The shaded regions designate the central 90\% interval across all
    times and datasets.
    }
  \label{fig:EMF_ali}
\end{figure}

Moving on to the EMF, 
the nonlinear closure outperforms the 
traditional $\alpha-\beta-\gamma$ closure  
in almost all datasets, maintaining a constant coefficient
with a median correlation of 0.79 (\figref{fig:EMF_corr}).
The traditional closure exhibits consistently weaker correlations,
despite the increased flexibility of three free coefficients.
Only $\Cb$, related to the turbulent resistivity term in the EMF, 
	is approximately constant, whereas $\Ca$ and $\Cg$ 
fail to maintain steady values or consistent signature. 
The reason for the consistently better correlations of the nonlinear closure 
is hinted at in \figref{fig:EMF_ali}. 
This probability density plot of the local alignment between $\EMFdata$ and $\EMF$ 
demonstrates that the traditional closure is almost randomly aligned
(flat distribution) whereas the nonlinear closure approaches the desired
$\delta$-distribution at $0^\circ$.

\section{Conclusions  and outlook}
In summary, we have proposed a set of constant coefficient closures
for the SGS stress and EMF in the filtered MHD equations and conducted
\aprio tests.
The tests we performed do show that the new nonlinear closures perform significantly better 
than traditional, phenomenological
closures with respect to both structural and functional diagnostics. The tests consist of
filtering \Enzo and \FLASH simulations of homogeneous, isotropic turbulence
and comparing the resulting SGS terms to their respective closures
(dependent only on the filtered fields).
All quantities are compared via their contributions to the SGS energy 
flux $\Sigma^{\Box}$, 
where the closure coefficients are computed by individual least-square
fitting.
In addition, the alignment for the EMF vector is investigated.
All new coefficients correlate well with the data.
They are constant over time and as a direct consequence the proposed closures
may be implemented in large-eddy simulations without the need for a computationally expensive
dynamical procedure which computes the coefficient values at run time.
In addition, the coefficients remain constant across simulation runs from two different codes and 
a wide range of plasma parameters, suggesting that the proposed closures capture
an underlying physical mechanism at work in highly compressible turbulent plasma flows.
Moreover, the new closures successfully represent the turbulent magnetic pressure,
reproduce the bi-directional energy cascade and are well aligned with the EMF.
We recognize the slightly lower correlation of the nonlinear closure in the
EMF than in the SGS stress counterpart, suggesting small room for improvement.

Nevertheless, the performance improvement over the traditional closures
already supports the implementation and validation of the new closures in an SGS model 
for large-eddy simulations of compressible turbulent plasma flows. 
These simulations would then allow us to infer the effect of the proposed model on the large scale flow in practice.
Potential applications include accretion disks \cite{Clark2011b}, star-forming magnetized clouds 
\cite{Greif2008, Greif2012a}
and plasmas on cosmological scales \cite{Agertz2013, Devriendt2010,Vazza2014,Schleicher2013a, Latif2013, Latif2014}.

\ack
PG worked on section \ref{sec:NonLinClosures} and conducted the \Enzo simulations.
DV worked on section \ref{sec:TradClosures}.
The analysis framework is a joint development by PG and DV.
  PG acknowledges financial support by the 
  \textit{International Max Planck Research School for Solar System 
  Science at the University of Göttingen}.
  DV, WS and DS acknowledge research funding by the 
  \textit{Deutsche Forschungsgemeinschaft (DFG)} under grant
  \textit{SFB 963/1, project A15}.
  CF thanks for funding provided by the ARC (grants DP130102078 and DP150104329).
  We gratefully acknowledge the J\"ulich Supercomputing Centre (grant hhd20), the Leibniz
  Rechenzentrum and the Gauss Centre for Supercomputing (grant pr32lo), the Partnership for
  Advanced Computing in Europe (PRACE grant pr98mu), and the Australian National Computing
  Infrastructure (grant ek9). The software used in this work was in part developed by the 
  DOE-supported Flash Center for Computational Science at the University of Chicago.
 
\section*{References}
\bibliographystyle{iopart-num}
\bibliography{MHDSGS_NJP}
\end{document}